\documentclass[12pt]{article}
\usepackage[cp1251]{inputenc}

\usepackage[english]{babel}

\usepackage{amsmath}
\usepackage{amssymb,amsthm}

\usepackage{verbatim}

\usepackage{graphicx}

\numberwithin{equation}{section}

\newcounter{Draft}
\usepackage[usenames]{color}
\usepackage{colortbl}
\usepackage{ifthen}

\usepackage[all]{xy}

\usepackage{amscd}

\renewcommand{\baselinestretch}{1.5}

\begin{document}

\begin{center}
{\Large A HelloWord \textsc{Bib}\negthinspace\TeX~stile file
.\textbf{bst}}\\
\vskip 5mm {\large Makar Plakhotnyk}\\
\vskip 5mm
S\~ao Paulo University, Brazil.\\
 mail:\, makar.plakhotnyk@gmail.com
\end{center}

\tableofcontents

\newpage
\section{Short motivation}

I was preparing an article to a mathematical journal. It uses
\textsc{Bib}\negthinspace\TeX, but there were no the necessary
file of bibliographical styles on its internet cite. Precisely,
there were some, but all of them generated not the same style of
references as in the articles of recent issues of the journal. I
have not understood, how the authors of these articles prepared
there files, but I had not imagined nothing better than to begin
to change one of the bibliography styles, which I have downloaded
from the cite of the journal.

I was doing my changes without any additional literature, just
with the help of my intuition, through trials and errors. After
the end of the work, which, frankly said, was not done properly, I
tried to understand, how can I prepare my own bibliography style:
how complicated it is (I was sure that it should not be very
complicated), where can I read about it etc.

I have found the program makebst (but not used it). I have read
that it asks a ``long list of questions'' and then produces the
bst-file (file with the style of bibliography). I have not even
downloaded and tried to use this program ``makebst'' because of I
had understood already that a bibliography style (\textbf{.bst}
-file) is a small program to produce the bibliography. It seamed
to me very strange to have a program (makebst), which asks some
questions and produces another program.

I have found~\cite{Markey} and~\cite{Patash}. I have found enough
information in these books (precisely in the first one, which is
bigger and contains more explanations), but I have spent a
mountain of time to prepare my bibliography style, because these
books are written as ``reference book'' with minimum of examples
and without detailed explanation ``what to start with''.

The sing, which seamed a very strange for me in the book of Oren
Patashnik is his suggestion that the best for the create a new
style is to start with an existing style that is close to yours,
and then modify it. Clearly, if somebody works in the editorial
staff of a huge journal, does not know how to create a
bibliographical style but, in the same time, needs to create
it,... hm... then indeed the best way is to modify (or, better,
not to modify) some of the existing styles. But if a person
studies to create styles, then (I think) the best way is to start
with a small ``Hello-Word'' style and then to improve it.

Thus I want to write about the preparation a new bibliography
style in the manner of writing an introduction to a new language
of programming. I want to start with a short ``HelloWord'' style
and then explain how one can improve it to something, which ``can
be used in the practical life''. I am not a professional in
latex-style creating, but this short book is just my own
experience of the preparation of a \textsc{Bib}\negthinspace\TeX
style-file.

\newpage
\section{The simplest examples}

\subsection{The use of bibliography citing without \textsc{Bib}\negthinspace\TeX}

Look at the file \textbf{test.tex} as at Figure~\ref{fig-1}. After
the execution (2 times\footnote{I will explain in
Section~\ref{sect-2}, why do we have to execute this program 2
times, but not 1.}) of \textbf{latex.exe test} it will be
generated the file \textbf{test.dvi} as at Figure~\ref{fig-2}.

The convenience here is that we can add or rearrange the
bibliographical items, and in this case the commands
\texttt{\textbackslash cite\{Ulam-1964\}} and
\texttt{\textbackslash cite\{Poincare\}} will generate the correct
numbers in brackets. For example, if we will add the first
bibliographical reference, i.e. will transform the bibliography as
at Figure~\ref{fig-3}, then the same command
\texttt{\textbackslash cite\{Ulam-1964\}} will generate~``[3]''
(see Figure~\ref{fig-4}).

\renewcommand{\baselinestretch}{1}
\begin{figure}[ht]
\begin{center}
\begin{tabular}{|p{0.95\linewidth}|} \hline
\texttt{\textbackslash documentclass\{article\} }

\noindent \texttt{\textbackslash begin\{document\}}\vskip 2mm

 \noindent \texttt{We cite the article\textasciitilde\textbackslash
 cite\{Ulam-1964\} by S. Ulam and the book\textasciitilde\textbackslash
cite\{Poincare\} by H. Poincar\textbackslash'e. Then we cite the
article\textasciitilde\textbackslash cite\{Ulam-1964\} by S. Ulam
again.}   \vskip 2mm

\noindent\texttt{\textbackslash begin\{thebibliography\}\{10\}} \\

\noindent\texttt{\textbackslash bibitem\{Poincare\} H. Poincar\'e,
\{\textbackslash it Les m\'ethods nouvelles de la m\'ecanique
c\'eleste\}, Paris: Gauthier-Villars, 1892.} \vskip 2mm

\noindent\texttt{\textbackslash bibitem\{Ulam-1964\} P. Stein and
S. Ulam, ``Non-linear transformation studies on electronic
computers'', \{\textbackslash it Rozprawy Mat.\}, Vol.
\{\textbackslash bf 39\}, pp.~1-66, 1964.} \\

\noindent\texttt{\textbackslash end\{thebibliography\}} \vskip 2mm

\noindent \texttt{\textbackslash end\{document\}}\\

\hline
\end{tabular}\end{center}\caption{test.tex}\label{fig-1}
\end{figure}
\renewcommand{\baselinestretch}{1.5}

\begin{figure}[ht]
\begin{center}
\begin{tabular}{|p{0.95\linewidth}|} \hline
We cite the article~[2] by S. Ulam and the book [1] by H.
Poincar\'e. Then we cite the article [2] by S. Ulam again.

\\

\hline
\end{tabular}\end{center}\caption{test.dvi}\label{fig-2}
\end{figure}

\renewcommand{\baselinestretch}{1}
\begin{figure}[ht]
\begin{center}
\begin{tabular}{|p{0.95\linewidth}|} \hline
\noindent\texttt{\textbackslash begin\{thebibliography\}\{10\}}\\

\noindent\texttt{\textbackslash bibitem\{Simson-2015\} Simson D.,
Tame-wild dichotomy of Birkhoff type problems for nilpotent linear
operators, J.  Algebra, Vol. 424 (2015) 254--293.}\\

\noindent\texttt{\textbackslash bibitem\{Poincare\} H. Poincar\'e,
\{\textbackslash it Les m\'ethods nouvelles de la m\'ecanique
c\'eleste\}, Paris: Gauthier-Villars, 1892.} \\

\noindent\texttt{\textbackslash bibitem\{Ulam-1964\} P. Stein and
S. Ulam, ``Non-linear transformation studies on electronic
computers'', \{\textbackslash it Rozprawy Mat.\}, Vol.
\{\textbackslash bf 39\}, pp.~1-66, 1964.} \\

\noindent\texttt{\textbackslash end\{thebibliography\}}\\

\hline
\end{tabular}\end{center}\caption{test-a.tex}\label{fig-3}
\end{figure}
\renewcommand{\baselinestretch}{1.5}

\begin{figure}[ht]
\begin{center}
\begin{tabular}{|p{0.95\linewidth}|} \hline
We cite the article~[3] by S. Ulam and the book [3] by H.
Poincar\'e. Then we cite the article [3] by S. Ulam again.

\\

\hline
\end{tabular}\end{center}\caption{test-a.dvi}\label{fig-4}
\end{figure}

\subsection{The use of \textsc{Bib}\negthinspace\TeX}

One of the most simple motivations to use the
\textsc{Bib}\negthinspace\TeX~is to standardize the presentation
of the bibliographical items in the bibliography. For example, it
should be clear that articles [1] and [3] at Figure~\ref{fig-4}
are described in different styles. For instance, in [1] the second
name comes before the first one, but in [3] is conversely. The
name of the journal is written by italic in [3], but in [1] is
not. The same thing is about the letters ``pp.'' before the pages
of article, bold style of the number of the volume, manner to
write the year.

Let us imagine, that we are a editorial staff of a scientific
journal. Clearly, we have some concrete rules for the manner of
presentation of an item of the bibliography of an article (i.e.
which should be bold, which should be italic, where should be
parenthesis, where slashes etc.). One way to achieve this is all
the articles of our journal is to give to all the authors an
``example'' (for instance, Figure~\ref{fig-2}) and to ask ``to
prepare the bibliography according to example''. Another way is to
ask them to prepare the data about their bibliographical items
separately and then we will use some program to prepare the
bibliography in the manner we need.

\renewcommand{\baselinestretch}{1}
\begin{figure}[ht]
\begin{center}
\begin{tabular}{|p{0.95\linewidth}|} \hline

\noindent\texttt{@article\{Ulam-1964,}\\
\texttt{\hskip 1cm author = "Stein P. R. and  Ulam S. M.",}\\
\texttt{\hskip 1cm title = "Non-linear transformation studies on
electronic computers",}\\
\texttt{\hskip 1cm journal = "Rozprawy Mat.",}\\
\texttt{\hskip 1cm year = "1964",}\\
\texttt{\hskip 1cm volume = "39",}\\
\texttt{\hskip 1cm pages = "1-66"\}}\\

\noindent\texttt{@book\{Poincare,}\\
\texttt{\hskip 1cm author = "H. Poincar\'e",}\\
\texttt{\hskip 1cm title = "Les m\'ethods nouvelles de la
m\'ecanique c\'eleste",}\\
\texttt{\hskip 1cm year = "1892",}\\
\texttt{\hskip 1cm publisher = "Gauthier-Villars",}\\
\texttt{\hskip 1cm address   = "Paris"\}}\\
\hline

\end{tabular}\end{center}\caption{my.bib}\label{fig-5}
\end{figure}

\begin{figure}[ht]
\begin{center}
\begin{tabular}{|p{0.95\linewidth}|} \hline

\noindent\texttt{\textbackslash documentclass\{article\}}\\
\texttt{\textbackslash begin\{document\}}\\
\texttt{We cite the article \textbackslash cite\{Ulam-1964\} and
the book \textbackslash cite\{Poincare\}.\}}\\

\texttt{\textbackslash bibliographystyle\{plain\}}\\
\texttt{\textbackslash bibliography\{my\}}\\
\texttt{\textbackslash end\{document\}}\\

\hline
\end{tabular}\end{center}
\caption{test2.tex}\label{fig-6}
\end{figure}
\renewcommand{\baselinestretch}{1.5}

The bibliography from Figure~\ref{fig-1} (the lines between
\noindent\texttt{\textbackslash begin\{thebibliography\}\{10\}}
and \noindent\texttt{\textbackslash end\{thebibliography\}})
should be redone as at Figure~\ref{fig-5} and saved in the
separate file with extension \textbf{.bib} (call it
\textbf{my.bib}). Also redo the \textbf{.tex}-file as at
Figure~\ref{fig-6} and name it \textbf{test2.tex}.

The line \texttt{\textbackslash bibliography\{my\}} in
\textbf{test2.tex} means that the bibliography is saved in the
same directory as the current file (\textbf{test2.tex}) and this
bibliography is named \textbf{my.bib}.

The line \texttt{\textbackslash bibliographystyle\{plain\}} means
that ``plain'' is the name of the style of bibliography, which we
will use (this is one of standard bibliography styles ant it is
generally distributed with \textsc{Bib}\negthinspace\TeX). The
bibliography stile \underline{\emph{plain}} is distributed
together with the \textsc{Bib}\negthinspace\TeX. There is a file

\centerline{\textbackslash bibtex\textbackslash bst\textbackslash
base\textbackslash plain.bst} \noindent (in the folder, where
\TeX~is installed), and this file provides the style
``plain''\footnote{also there are ``another styles'' in the
mentioned folder. The extensions of all of them are .\textbf{bst}
and it is an extension of a file of
\textsc{Bib}\negthinspace\TeX-style.}.

To create the file \textbf{test2.dvi} we need to:

1. execute \textbf{latex.exe test2} (one time),

2. execute \textbf{bibtex.exe test2} (one time),

3. execute \textbf{latex.exe test2} (two times).

\begin{figure}[ht]
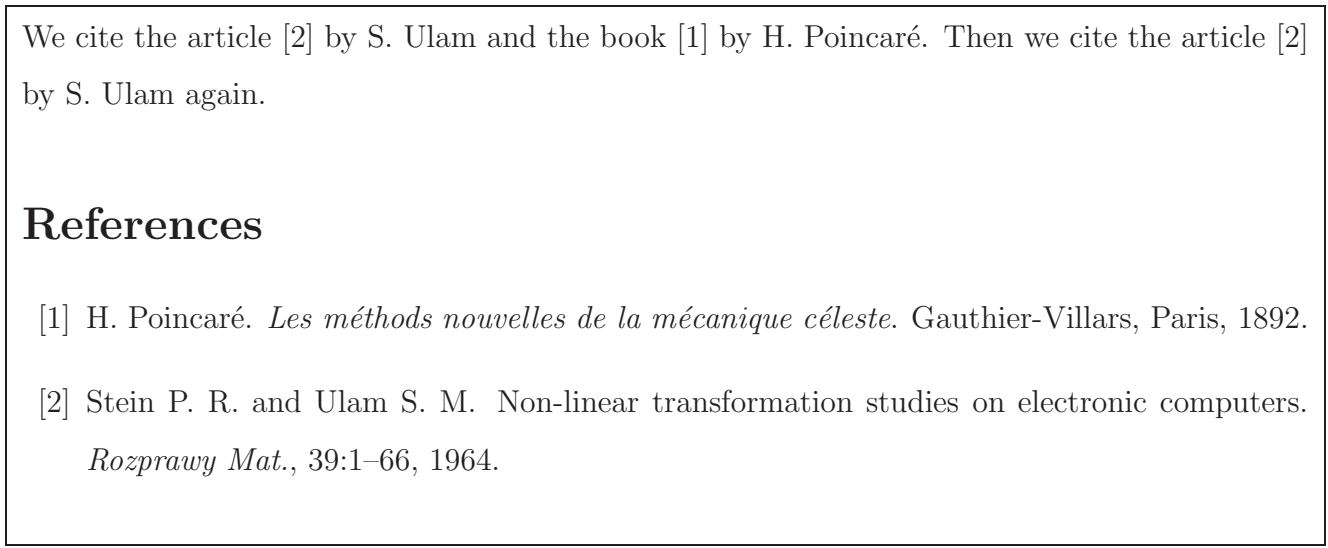

\begin{center}
\begin{tabular}{|p{0.95\linewidth}|} \hline
We cite the article~[2] by S. Ulam and the book [1] by H.
Poincar\'e. Then we cite the article [2] by S. Ulam again.

\\

\hline
\end{tabular}\end{center}\caption{test2.dvi}\label{fig-10}
\end{figure}

After these procedures our final test looks as at
Figure~\ref{fig-10}. Remark, that this is not exactly the same, as
we had at Figure~\ref{fig-2}. In fact, the file \textbf{test2.bbl}
was created by the program \textbf{bibtex.exe} and it contains the
bibliography, which is very similar to one from
Figure~\ref{fig-1}.

\renewcommand{\baselinestretch}{1}

\begin{figure}[ht]
\begin{center}
\begin{tabular}{|p{0.95\linewidth}|} \hline

\noindent\texttt{ENTRY}\\
\texttt{\hskip 1cm \{ author}\\
\texttt{\hskip 1cm \}\{\}\{\}}\\

\noindent\texttt{FUNCTION \{output.bibitem\} \{}\\
\texttt{\hskip 1cm "\textbackslash bibitem\{" write\$}\\
\texttt{\hskip 1cm cite\$ write\$}\\
\texttt{\hskip 1cm "\}" write\$ newline\$\} }\\

\noindent\texttt{FUNCTION \{article\}\{}\\
\texttt{\hskip 1cm output.bibitem}\\
\texttt{\hskip 1cm   author write\$}\\
\texttt{\hskip 1cm " (article)" write\$}\\
\texttt{\hskip 1cm newline\$ \}}\\

\noindent\texttt{FUNCTION \{book\}\{}\\
\texttt{\hskip 1cm output.bibitem}\\
\texttt{\hskip 1cm author write\$}\\
\texttt{\hskip 1cm " (book)" write\$}\\
\texttt{\hskip 1cm newline\$ \}}\\
\texttt{READ}\\

\noindent\texttt{FUNCTION \{begin.bib\}\{}\\
\texttt{\hskip 1cm "\textbackslash begin\{thebibliography\}\{10\}"
write\$}\\
\texttt{\hskip 1cm newline\$ newline\$ \}}\\

\noindent\texttt{EXECUTE \{begin.bib\}}\\
\texttt{ITERATE \{call.type\$\}}\\
\texttt{FUNCTION \{end.bib\}\{}\\
\texttt{\hskip 1cm "\textbackslash end\{thebibliography\}"
write\$ \}}\\
\texttt{EXECUTE \{end.bib\}}\\

\hline
\end{tabular}\end{center}
\caption{helloword.bst}\label{fig-9}
\end{figure}

\renewcommand{\baselinestretch}{1.5}

\subsection{The Hello-Word \textsc{Bib}\negthinspace\TeX-file}

The style, which we will consider the simplest (hello-word) style
is given at Figure~\ref{fig-9}. To use this file we save our
\textbf{test2.tex} as \textbf{test3.tex} and just change the
parameter plain of the command \emph{\textbackslash
bibliographystyle} into \emph{helloword}. Save the ``script'' from
Figure~\ref{fig-9} in the same folder, where \textbf{test3.tex}
is, and let's name this style-file \textbf{helloword.bst}.

After the execution of \textbf{latex.exe test3} and then
\textbf{bibtex.exe test3} the file \textbf{test3.bbl} (see
Figure~\ref{fig-11}) is generated. The final file
\textbf{test3.dvi} is given at Figure~\ref{fig-12}.

Clearly, the obtained file \textbf{test3.dvi} corresponds the
bibliography-file \textbf{test3.bbl}. This \textbf{test3.bbl} was
created ``by the algorithm'', which is there in
\textbf{helloword.bst}.

I shall not give the detailed explanations about ``how
\textbf{helloword.bst} works'', but we will just briefly look
through \textbf{helloword.bst} and compare it with
\textbf{test3.bbl}.

\renewcommand{\baselinestretch}{1}
\begin{figure}[ht]
\begin{center}
\begin{tabular}{|p{0.95\linewidth}|} \hline

\noindent\texttt{\textbackslash
begin\{thebibliography\}\{10\}}\vskip 5mm

\noindent\texttt{\textbackslash bibitem\{Ulam-1964\}}\\
\texttt{Stein P. R. and Ulam S. M. (article)}\\
\noindent\texttt{\textbackslash bibitem\{Poincare\}}\\
\texttt{H. Poincar\'e (book)}\\
\noindent\texttt{\textbackslash end\{thebibliography\}}\\

\hline
\end{tabular}\end{center}\caption{test3.bbl}\label{fig-11}
\end{figure}

\begin{figure}[ht]
\begin{center}
\begin{tabular}{|p{0.95\linewidth}|} \hline
We cite the article~[1] by S. Ulam and the book [2] by H.
Poincar\'e. Then we cite the article [1] by S. Ulam again.

\\

\hline
\end{tabular}\end{center}\caption{test3.dvi}\label{fig-12}
\end{figure}

\renewcommand{\baselinestretch}{1.5}

The second line \texttt{\{ author \}} in \textbf{helloword} means
that the field ``author'' will be accessible to our program (thus,
another fields from \textbf{my.bib}, i.e. ``title'', ``journal'',
``year'', ``volume'', ``publisher'' and ``address'' will not. In
other words, our style-file (our ``algorithm'') will not be able
to use these fields. To make one (or more than one) of these
parameters accessible we had to write for example:

\noindent\texttt{ENTRY}\\
\hskip 1cm \texttt{ \{ author}\\
\hskip 1cm \texttt{ title}\\
\hskip 1cm \texttt{ journal}\\
\texttt{\hskip 1cm \}\{\}\{\}}\\

After the block called ENTRY, we see functions
\emph{output.bibitem}, \emph{article} and \emph{book}. The
functions will be executed (if will be) a bit later by the
correspond commands (and we will see these commands).

The word ``READ'' is the command, which tells
\textsc{Bib}\negthinspace\TeX ``to read our file
\textbf{test3.tex}''\footnote{indeed this command says to read the
file \textbf{test3.aux}, which we will discuss a bit later, but
for the first time we can imagine that the file \textbf{test3.tex}
is read.}

After the command \texttt{READ} we have the function
\texttt{begin.bib}, which will be executed later, in the same
manner as another functions.

The next command \texttt{EXECUTE \{ begin.bib \}} executes the
function \emph{begin.bib}, which should be described later,
because \textsc{Bib}\negthinspace\TeX-file reads the
.\textbf{bst}-file ``line by line'' and whence ``does not know'',
what is there later; thus, in the moment of the command to execute
the function \emph{begin.bib}, this function should be already
described.

Since the we have already seen the function \emph{begin.bib} early
in our file, let us come there and understand, what it does. It
does nothing more that to write the line ``\textbackslash
begin\{thebibliography\}\{10\}'' onto the file \textbf{test3.bbl},
then write two times the symbols or the begin of the next line
(which is clear from the file \textbf{test3.bbl}, which, in fact,
appears).

After the function \emph{begin.bib} finished its execution, we
come back to the line, where it was called and look at the next
line.

The next command \texttt{ITERATE \{call.type\$\}} does the
following:

1. looks through our file \textbf{test3.tex}\footnote{indeed,
through \textbf{test3.aux}, but we will talk about this later.}.

2. takes every command \textbackslash ref\{...\} (in the order or
the appearance in the file); precisely, in our file it is
\textbackslash cite\{Ulam-1964\}, then \textbackslash
cite\{Poincare\} and then \textbackslash cite\{Ulam-1964\} ones
more.

3. Find the description of the reference \textbackslash
cite\{Ulam-1964\} in the file \textbf{my.bib}. This description
starts with ``\emph{@article\{Ulam-1964},''. Thus, the function
\emph{article} will be executed.

4. The same thing will happen with \textbackslash
cite\{Poincare\}. First the line ``\emph{@book\{Poincare,}'' will
be found in \textbf{my.bib}, and the the function called
\emph{book} will be executed.

5. Since the third citation \textbackslash cite\{Ulam-1964\}
corresponds to the reference, which was prepared for the
\textbf{test3.bbl} already, then happens nothing.\\

This idea (with different functions ``article'' and ``book'') is
realized to make possible to prepare the bibliographical
description of  ``articles'' and ``books'' in different manner.
For example, every article has journal (where it was published),
i.e. during the creation of the description of the article the
field ``journal'' should be used. Nevertheless, a book does not
have such parameter as journal.

In any way, now we will look inside the function \emph{article}
(since it was executed first) and then inside the function
\emph{book}.

At the very beginning the function \emph{article} calls one more
function: \emph{output.bibitem}. This is written just by the first
line of \emph{article}. The function \emph{output.bibitem} works
in the same way, as we have already seen in \emph{begin.bib}.
First write into the file \textbf{test3.bbl} the sequence of
symbols ``\textbackslash bibitem\{''. The command ``cite\$'' takes
the name of the reference (in or case it is ``Ulam-1964'') and
writes in to \textbf{test3.bbl}. Then close the bracket ``\}'' in
the \textbf{test3.bbl} and writes the symbol of the new line.
Thus, in our case the line \centerline{\textbackslash
bibitem\{Ulam-1964\}} \noindent appears. After this
\emph{output.bibitem} is over and program goes back to
\emph{article}. The command \texttt{author write\$} writes into
\textbf{test3.bbl} the text from the field ``author'' of the
reference ``Ulam-1964''. The next command \texttt{" (article)"
write\$} writes `` (article)''. We do it to distinguish the
function \emph{article} and the function \emph{book} in the final
file \textbf{test3.bbl}.

The function \emph{book} works absolutely analogically to
\emph{article}.

When the functions \emph{article} and \emph{book} are finished,
the program comes to the line, which is next after \texttt{ITERATE
\{call.type\$\}}. After this we see the description of the
function \emph{end.bib}. And then our program finishes with the
execution of \emph{end.bib}, i.e. with writing the text

\centerline{\textbackslash end\{thebibliography\}} \noindent into
\textbf{test3.bbl}.

\newpage
\section{The .aux-file, or how does \LaTeX creates
the bibliographical citations}\label{sect-2}

\subsection{The case, where \textsc{Bib}\negthinspace\TeX is not used}

Lets come back to the file \textbf{test.tex} (see
Figure~\ref{fig-1}) and suppose that there are no any other file
in this folder.

During the first run of \textbf{latex.exe test} it creates three
files:

1. test.log (the comments during the translation),

2. test.aux (information about the bibliography),

3. test.dvi (the result of the translation).\\

The \textbf{test.aux} file after the first run of
\textbf{latex.exe test} is as at the Figure~\ref{fig-7}. This file
is easy to understand. The first line \texttt{\textbackslash
relax} is not important for us\footnote{In fact, this is a command
which says to \LaTeX to ``stop the execution of any function if it
is executed''.}. When \LaTeX~meets the citation \textbackslash
ref\{...\} in the .\textbf{tex}-file, it creates a line
\textbackslash citation\{...\} in the .\textbf{aux}-file.

\begin{figure}[ht]
\begin{center}
\begin{tabular}{|p{0.95\linewidth}|} \hline
\noindent\texttt{\textbackslash relax}\\
\texttt{\textbackslash citation\{Ulam-1964\}}\\
\texttt{\textbackslash citation\{Poincare\}}\\
\texttt{\textbackslash citation\{Ulam-1964\}}\\
\texttt{\textbackslash bibcite\{Poincare\}\{1\}}\\
\texttt{\textbackslash bibcite\{Ulam-1964\}\{2\}}\\

\hline
\end{tabular}\end{center}\caption{test.aux}\label{fig-7}
\end{figure}

When \LaTeX meets the bibliography item in the .\textbf{tex}-file,
it creates a line \textbackslash bibitem\{...\}, it creates the
line \textbackslash bibcite\{...\} in the .\textbf{aux}-file.

Nevertheless, the final file \textbf{tes.dvi} does not have
references after the first execution of \textbf{latex.exe test}
(see Figure~\ref{fig-8}).

\begin{figure}[ht]
\begin{center}
\begin{tabular}{|p{0.95\linewidth}|} \hline
We cite the article~[?] by S. Ulam and the book [?] by H.
Poincar\'e. Then we cite the article [?] by S. Ulam again.

\\

\hline
\end{tabular}\end{center}\caption{test.dvi}\label{fig-8}
\end{figure}

We can found a hint about these [?]s in the file
\textbf{test.log}. It contains a line ``No file test.aux'', which,
clearly, mans that \textbf{latex.exe test} tried to use
\textbf{test.aux}, but this file was not created this moment.
Remark, that the messages

\centerline{LaTeX Warning: Citation `Ulam-1964' on page 1
undefined on input line ...}

\centerline{LaTeX Warning: Citation `Poincare' on page 1 undefined
on input line ...}

\noindent appear after the message about non-existence of
\textbf{test.aux}. We can explain it as follows. Before the begin
of the translation the file \textbf{test.tex}, the \LaTeX~opens
(tries to open) the \textbf{test.aux}. It is successes with this,
it reads all the lines \textbackslash bibcite\{...\}, which give
the correspondence with bibliographical references (``Ulam-1964''
and ``Poincare'' in our case) and numbers (2 and 1 respectively in
our case). If the file \textbf{test.aux} is read (precisely if it
exists) then the relations, obtained from it, are used to
construct the references ``[1]'' and ``[2]'' instead of ``[?]''.

In any way, after the second execution of \textbf{latex.exe test},
the message \texttt{No file 1.aux} disappears from
\textbf{test.log} and we obtain the final result
\textbf{test.dvi}, given at Figure~\ref{fig-2}.

We can present here one more experiment, which will clarify our
understanding of the creation of the bibliography by \LaTeX. Let
us delete the file \textbf{test.aux} and execute \textbf{latex.exe
test}. Then enter the \textbf{test.aux} and change the last two
lines to\\
\texttt{\textbackslash bibcite\{Poincare\}\{10\}}\\
\texttt{\textbackslash bibcite\{Ulam-1964\}\{25\}}\\

\begin{figure}[ht]
\begin{center}
\begin{tabular}{|p{0.95\linewidth}|} \hline
We cite the article~[25] by S. Ulam and the book [10] by H.
Poincar\'e. Then we cite the article [25] by S. Ulam again.

\\

\hline
\end{tabular}\end{center}\caption{test.dvi}\label{fig-13}
\end{figure}

After the second execution of \textbf{latex.exe test} we will
obtain the file \textbf{test.dvi} as at Figure~\ref{fig-13}. The
message

\centerline{Warning: Label(s) may have changed. Rerun to get
cross-references right.} \noindent will appear in
\textbf{test.log}. The file \textbf{test.aux} will return to its
``previous'' condition, i.e. as at Figure~\ref{fig-7}.

After one more execution of \textbf{latex.exe test}, the
\textbf{test.dvi} also will come back to the form, which is
presented at Figure~\ref{fig-2}.

\subsection{The use of \textsc{Bib}\negthinspace\TeX}

When we execute \textbf{latex.exe test2} for the file
\textbf{test2.tex} (see Figure~\ref{fig-6}), then (in the
assumption, that there is no \textbf{test2.bbl} in the folder) the
file \textbf{test2.aux} as at Figure~\ref{fig-20} is generated.

\renewcommand{\baselinestretch}{1}
\begin{figure}[ht]
\begin{center}
\begin{tabular}{|p{0.95\linewidth}|} \hline
\noindent\texttt{\textbackslash relax}\\
\texttt{\textbackslash citation\{Ulam-1964\}}\\
\texttt{\textbackslash citation\{Poincare\}}\\
\texttt{\textbackslash citation\{Ulam-1964\}}\\
\texttt{\textbackslash bibstyle\{plain\}}\\
\texttt{\textbackslash bibdata\{my\}}\\

\hline
\end{tabular}\end{center}\caption{test2.aux}\label{fig-20}
\end{figure}
\renewcommand{\baselinestretch}{1.5}

Now, when we will execute \textbf{bibtex.exe test2}, this
\textbf{test2.aux} will be analyzed at the very beginning (not
\textbf{test2.tex}). Clearly, the style \emph{plain} will be used
after (because exactly this style is mentioned in the file) the
bibliography will be taken from \textbf{my.bib}. After all of this
the file \textbf{test2.bbl} will be generated.

When we already have \textbf{test2.bbl}, we may execute
\textbf{latex.exe test2} and it will add to \textbf{test2.aux} two
lines:\\
\renewcommand{\baselinestretch}{1}
\noindent \texttt{\textbackslash bibcite\{Poincare\}\{1\}}\\
\texttt{\textbackslash bibcite\{Ulam-1964\}\{2\}}\\
\renewcommand{\baselinestretch}{1.5}
Nevertheless, the bibliographical links will not be generated in
\textbf{test2.dvi} (after the first execution of \textbf{latex.exe
test}). The explanation of this is the same why they were not
generated after the \underline{first} execution of
\textbf{latex.exe test} in previous subsection.

\newpage
\section{Polish notation and some explanations
about the style of programming in .bst-file}

The polish notation is the tradition to write formulas (functions
with arguments) such that first we write arguments and then the
name of the function. For example, to write $$ 1; 2; +$$ instead
of
$$1 + 2.$$
Moreover, every function (including the algebraic operations),
which has input arguments and which has output, works with a
\textbf{\underline{stack}} ``first in -- last out''.

For example, when style the file \textbf{test.bst} (from
Figure~\ref{fig-9}) was executed:

1. the line \texttt{"\textbackslash bibitem\{" write\$} meant to
write ``\textbackslash bibitem\{'' to the stack; then execute the
function \emph{write\$}, which reads one parameter from the stack
and writes it to the file \textbf{test.bbl} (indeed, not exactly
to file, but to the ``buffer'' of \textbf{test.bbl}) After all the
writing to the buffer of \textbf{test.bbl} we need to use the
function \emph{newline\$}, which writes everything from the buffer
to the file and then writes a symbols of the end of a line into
the file.

2. the line \texttt{author write\$} worked similarly. First write
to the stack the field ``author'' and then write one string from
the stack to the buffer of \textbf{test.bbl}.

There are some natural rules of using this stack. For example:

1. Assume that at the moment of the start of the execution of the
.\textbf{bst}-file the stack is empty. This means that if we use a
function, which reads something from the stack, then we should be
sure that this ``something'' is there in written to the stack
earlier.

2. Variables has types. The types are only integer and string. If
some function has integer argument, then correspond its argument
in the stack should be integer, but not string.

3. At the end of the execution of the .\textbf{bst}-file the stack
should be empty.

We will not give now the list of all the function (this list with
the descriptions is there, for example, in the book by Nicolas
Markey), but we will give some remarks just to understand, what
the functions from that list ``can do'':

1. There is an operator ``:='', which puts the value from the
stack to the variable and should be used as \texttt{'var :=} (i.e.
with apostrophe) to assign the value from the stack to the
variable \texttt{var};

2. Arithmetical operations are functions;

3. there is possibility to use cycles (function \emph{while}) and
to verify conditions (function \emph{if});

There are some other functions, but all of them are not important
for us just now.

\newpage
\section{Ordering of references with the use of
\textsc{Bib}\negthinspace\TeX-style}

As far as we has understood from the study of the our hello-word
style, that the simplest way to order the bibliography is by the
appearance in the \textbf{.tex}-file, or precisely (which, in
fact, is the same) in the \textbf{.aux}-file. The need to order
the bibliography in any other way (for example, by author's names)
is frequent and natural. It can be easily realized.

Let us take our style \textbf{test.bst} (from Figure~\ref{fig-9})
and add there the fragment, which is there at Figure~\ref{fig-14}.
Put this fragment after the command \emph{read} and before the
function \emph{begin.bib}.

\renewcommand{\baselinestretch}{1}
\begin{figure}[ht]
\begin{center}
\begin{tabular}{|p{0.95\linewidth}|} \hline

\noindent\texttt{FUNCTION \{bib.sort.order\}\{}\\
\texttt{\hskip 1cm author 'sort.key\$ := \}} \vskip 5mm
\noindent\texttt{ITERATE \{bib.sort.order\}}\\
\noindent\texttt{\{SORT\}}\\

\hline
\end{tabular}\end{center}\caption{}\label{fig-14}
\end{figure}
\renewcommand{\baselinestretch}{1.5}

Each bibliographical item has the string variable, called
\emph{sort.key}. The command SORT sorts the bibliographical items
according to this variable. The command $$\noindent\texttt{ITERATE
\{bib.sort.order\}}$$ executes our function \emph{bib.sort.order},
which sets the value of the field ``author'' to the variable
sort.key.

Thus, after such modification of the \textbf{.bst}-file we will
have our bibliography, ordered by author. From another hand,
attentive reader must notice, that ``life is a bit more
complicated'', because in our example we had authors ``H.
Poincar\'e'' and ``P. Stein and S. Ulam''. If they will be sorted
as we have done, then (also in the case o huge list of
bibliography) the first item will appear between items, starting
with letter ``H'' (H. Poincar\'e), and the same thing about ``P.
Stein and S. Ulam''. Nevertheless, it is natural to want to sort
bibliography by second names, but not complete names. We will
discuss this problem a bit later in the section~\ref{sect-3}.

\newpage
\section{Function if\$}

Clearly, the bifurcations (like ``if'') and cycles (like
``while'') is the basis of any language of programming. From
another hand, the \textbf{.bst}-file is not a pure program, but a
list of commands, how to create a \textbf{.bib}-file.

Let us see an example, where bifurcation is necessary.

\centerline{Wether the article has number?}

\noindent  Some journals identify the article by: volume (one year
is one volume), number (starts from one each year), pages. Some
another journal's issues has no number (but has only volume).

Suppose that we have one more reference (see Figure~\ref{fig-16})
and have citing ``\textbackslash cite\{YangYu\}'' somewhere in the
\textbf{.tex}-file.

\renewcommand{\baselinestretch}{1}

\begin{figure}[ht]
\begin{center}
\begin{tabular}{|p{0.95\linewidth}|} \hline

\noindent\texttt{@article\{YangYu,}

\texttt{\hskip 1cm author = "Yang Tse-Chung and Yu Chia-Fu",}

\texttt{\hskip 1cm title = "Monomial, Gorenstein and Bass
orders",}

\texttt{\hskip 1cm journal = "J. Pure Appl. Algebra",}\\
\texttt{\hskip 1cm year = "2015",}\\
\texttt{\hskip 1cm volume = "219",}\\
\texttt{\hskip 1cm pages = "767-778",}\\
\texttt{\hskip 1cm number = "4"\}}\\

\hline

\end{tabular}\end{center}\caption{}\label{fig-16}
\end{figure}
\renewcommand{\baselinestretch}{1.5}

Let us add fields ``number'' and ``volume'' to our style and lets
modify the function \emph{article} as at Figure~\ref{fig-15}.

\renewcommand{\baselinestretch}{1}
\begin{figure}[ht]
\begin{center}
\begin{tabular}{|p{0.95\linewidth}|} \hline

\noindent\texttt{ENTRY}\\
\texttt{\hskip 1cm \{ author}\\
\texttt{\hskip 1cm \{ number}\\
\texttt{\hskip 1cm \{ volume}\\
\texttt{\hskip 1cm \}\{\}\{\}}\\
\noindent\texttt{$\vdots$}\\
\noindent\texttt{FUNCTION \{article\}\{}\\
\texttt{\hskip 1cm output.bibitem}\\
\texttt{\hskip 1cm author write\$}\\

\texttt{\hskip 1cm ", No. " write\$ number write\$}\\
\texttt{\hskip 1cm ", Vol. " write\$ volume write\$}\\
\texttt{\hskip 1cm newline\$ \}}\\
\noindent\texttt{$\vdots$}\\

\hline
\end{tabular}\end{center}
\caption{}\label{fig-15}
\end{figure}

\begin{figure}[ht]
\begin{center}
\begin{tabular}{|p{0.95\linewidth}|} \hline
\\

\hline
\end{tabular}\end{center}\caption{}\label{fig-17}
\end{figure}
\renewcommand{\baselinestretch}{1.5}

The references of our file will look as at Figure~\ref{fig-17}.
Clearly, it is bad, because the word ``No.~'' should not appear in
the second reference. Moreover, the \textbf{bibtex.exe} generates,
during the generation a \textbf{.bib}-file its own file with
comments \textbf{test3.blg} and the comment

\centerline{`number' is a missing field, not a string, for entry
Ulam-1964}

\noindent appeared there. The significance of this comment is
clear. Modify our function book ones more as at the
Figure~\ref{fig-18}.

\renewcommand{\baselinestretch}{1}
\begin{figure}[ht]
\begin{center}
\begin{tabular}{|p{0.95\linewidth}|} \hline

\noindent\texttt{$\vdots$}\\
\noindent\texttt{FUNCTION \{article\}\{}\\
\texttt{\hskip 1cm output.bibitem}\\
\texttt{\hskip 1cm author write\$}\\
\texttt{\hskip 1cm number empty\$}\\
\texttt{\hskip 1cm \{skip\$\}}\\
\texttt{\hskip 1cm \{", No. " write\$ number write\$\}}\\
\texttt{\hskip 1cm  if\$}\\
\texttt{\hskip 1cm ", Vol. " write\$ volume write\$}\\
\texttt{\hskip 1cm newline\$ \}}\\
\noindent\texttt{$\vdots$}\\

\hline
\end{tabular}\end{center}
\caption{}\label{fig-18}
\end{figure}
\renewcommand{\baselinestretch}{1.5}

Explain a bit the obtained function. The function \emph{if\$} has
three arguments (which are taken from the stack). The first one is
an integer (which is considered as true if it is strictly positive
and false otherwise). The second argument should be command (of
name of another function, or sequence of commands). We have
written in brackets the first argument of if\$:

\centerline{\texttt{\{skip\$\}}}

\noindent and the second its argument

\centerline{\texttt{\hskip 1cm ", Vol. " write\$ volume write\$}\
.}

The command \texttt{\{skip\$\}} means ``do nothing'' and one of
the reason of its existence is, possibly, the way to use the
function if\$, as we have to. The third parameter of \emph{if\$}
should be clear, since it already appeared in the previous version
of our function \emph{book}.

The function \texttt{\{empty\$\}} checks, wether its argument
(taken from the stack as an argument of any function) is an empty
string. If it is, the result (which is put to the stack) 1 is
returned and 0 otherwise. Thus, the line

\centerline{number empty\$}

\noindent means that we first put ``number'' into the stack (if a
reference do not have some [declared] field, this field is
considered to be an empty string); then \emph{empty\$} puts a
string from the stack and checks, wether it is empty.

\newpage
\section{Function format.name\$}\label{sect-3}

There are some situations, which we need to obtain the second
named from the list of named of the authors (for example, for
order bibliography by author's names).

Traditionally all the information about author's names
\textsc{Bib}\negthinspace\TeX takes from the field ``author'',
whence it will be considered strange, if our style will have
fields like

\centerline{``first.authors.first.name'',
``first.authors.second.name'', ``second.authors.first.name'',}

\centerline{``second.authors.second.name'',
``third.authors.first.name'', ``third.authors.second.name'' etc.}

This ``traditionally'' it quiet important, because the idea of
\textsc{Bib}\negthinspace\TeX~is to create the style, which
manipulates with the bibliographical data-base and thus, our own
style should not need to ask the author of the \textbf{.bib}-file
to redo it in some way to satisfy the needs of especially our
style \textbf{.bst}.

Thus, there are some rules to prepare the names of authors in the
field author:

1. Different authors should be separated by the word ``and'' onr
from another

2. Author's name has four parts: First name; Last name; part
``von'', and ``suffix''.

2. There are three forms to write each of the author:

2a. First von Last;

2b von Last, First;

2c. von Last, Suffix, First.

There are two functions to work with names: \emph{num.names\$} and
\emph{format.name\$}.

The first of them (\emph{num.names\$}) is easy: it takes (clearly
from the stack) a string and computes the number of words ``and''
in it. This is assumed to calculate the number of authors in the
list of author's names.

The function \emph{format.name\$} has three arguments: String (the
field ``author''), integer (the number of the author in the list
of authors), and String (the output format of the name, which we
want).

The use of the function \emph{format.name\$} is clearly explained
in the book by N. Markey.

The example of the ``output format of the name'' is

\centerline{``\{ff\}\{vv\}\{ll\}\{jj\}'',}

\noindent which means here to obtain the (complete) first name
(i.e. everything, which is the first name), then the complete
von-part, then the complete last name and then the complete
suffix. We could write

\centerline{``\{ff\}\{l.\}\{jj\}''}

\noindent to take only the first latter (s) of the last name (to
write only one ``l'' instead of ``ll'' and to put dot after the
last name part. We could not ask about some parts about the name,
for example (like we have ignored the ``von''-part in the last
example.\\

Clearly, the existence of all of these rules means that the author
of the \textbf{.bib}-file has to prepare the list of names
``correctly'', i.e. in such a way that functions
\emph{num.names\$} and \emph{format.name\$} could obtain the list
of named of authors and the part of each author correctly.
Otherwise, the incorrect \textbf{.bib}-file should be considered
as a mistake of its author, but not of author or a style. For
example, if there is a rule to write the name as

\centerline{First von Last,}

\noindent then the author name should be written as

\centerline{F. Riss}

\noindent not ``Riss F.''. Nevertheless, the same name can be
written as

\centerline{Riss, F.}

\noindent (with comma) according to the rule

\centerline{von Last, First.}

\noindent In this case, can not be written ``F., Riss''.

\renewcommand{\baselinestretch}{1}
\begin{figure}[ht]
\begin{center}
\begin{tabular}{|p{0.95\linewidth}|} \hline

\noindent\texttt{FUNCTION \{bib.sort.order\}\{}\\
\texttt{\hskip 1cm author \#1 "\{ll\}" format.name\$}\\
\texttt{\hskip 1cm author num.names\$ \#1 =}\\
\texttt{\hskip 1cm \{skip\$\}}\\
\texttt{\hskip 2cm \{ author \#2 "\{ll\}" format.name\$ *}\\
\texttt{\hskip 2cm author num.names\$ \#2 =}\\
\texttt{\hskip 2cm \{skip\$\}}\\
\texttt{\hskip 2cm \{author \#3 "\{ll\}" format.name\$ *\}}\\
\texttt{\hskip 2cm if\$\}}\\
\noindent\texttt{\hskip 1cm 'sort.key\$ := \}}\\

\hline
\end{tabular}\end{center}
\caption{}\label{fig-19}
\end{figure}
\renewcommand{\baselinestretch}{1.5}

For example, the simplest function bib.sort.order is given at
Figure~\ref{fig-19}. It sorts the list of references by author's
second (last) names, but not by the strings ``author''.

\newpage
\section{Conclusion}

A good bibliographical style has to be written only by a person,
who has a lot of experience. This experience is not an experience
of programming, but the experience of checking the correctness of
the bibliographical references. The language of \textbf{.bst}-file
is simple enough, but this language is nothing more than a tool to
prepare the bibliography, but the logics of the preparation
depends of the mind of the programmer, but not on the language.
For example, we have noticed that an article may have a number of
a journal, may not have. An article may a very long list of
authors (up to 30 in some articles on physics). If somebody does
not know this, then only experience may help. And what obvious
(for an experienced person) and important we have not noticed?

If a person prepares his(her) own book with 20-30 bibliographical
references, then to prepare an own \textbf{.bst}-file may be an
interesting exercise. But if somebody works in the additional
staff of a journal, which prepares 100 articles a month with 20-30
references each, then usage (and improve) of an own
\textbf{.bst}-file, starting with hello-word can lead to a huge
disadvantages of one of the issues of this journal. From another
hand, any programmer, who studies a new language, do not do study
during the participation in a huge project and do not prepare a
huge project during one night.

\bibliography{Bibtexarx}{}
\bibliographystyle{plain}

\end{document}